\documentclass[10pt,conference]{IEEEtran}
\IEEEoverridecommandlockouts
\usepackage{cite}
\usepackage{amsmath,amssymb,amsfonts}
\usepackage{algorithmic}
\usepackage{textcomp}
\usepackage[usenames,dvipsnames]{xcolor}
\usepackage{url}
\usepackage{graphicx}
\usepackage{tikz}
\usepackage{inconsolata}
\usetikzlibrary{arrows.meta, shapes, positioning, fit}
\usepackage{comment}
\usepackage[htt]{hyphenat}
\usepackage{booktabs}
\usepackage{listings}
\usepackage{comment}
\usepackage{minted}
\usepackage{enumitem}
\usepackage{subcaption}
\usepackage[hidelinks]{hyperref}
\PassOptionsToPackage{hyphens}{url}\usepackage{hyperref}
\hypersetup{
    colorlinks=true,
    citecolor=blue,
    urlcolor=black
}
\newcommand{\tool}{{\sc pytrim}}
\newcommand{\empirical}[1]{#1}
\newcommand{\point}[1]{\par\smallskip\noindent{\textbf{#1:}} }

\newcommand{\linesOfCode}{$\sim$4,500}
\newcommand{\openSourcePackages}{971}
\newcommand{\projectsWithBloat}{39}
\newcommand{\mergedPRs}{14}
\newcommand{\pendingPRs}{19}
\newcommand{\closedPRs}{6}

\begin{document}

\title{PyTrim: A Practical Tool for Reducing Python Dependency Bloat}

\author{
Konstantinos Karakatsanis$^1$, 
Georgios Alexopoulos$^2$, 
Ioannis Karyotakis$^1$, 
Foivos Timotheos Proestakis$^1$\\
Evangelos Talos$^1$,
Panos Louridas$^1$, 
Dimitris Mitropoulos$^{2}$\\
\thanks{
This work has received funding from the EU’s Horizon 2021 research and innovation programme under grant agreement no. 101070599 (SecOPERA). ©2025 IEEE. Author’s manuscript. License CC BY 4.0. Accepted in the Proceedings of the 40th IEEE/ACM International Conference on Automated Software Engineering (ASE), Seoul, Korea. DOI: 10.1109/ASE63991.2025.00377
}
$^1$Athens University of Economics and Business, 
$^2$University of Athens\\
\{karakatsanis, karyotakisg, proestakis, vtalos, louridas\}@aueb.gr, \{grgalex, dimitro\}@ba.uoa.gr
}

\maketitle

\begin{abstract}
Dependency bloat is a persistent challenge in Python projects,
which increases maintenance costs and security risks.
While numerous tools exist 
for detecting unused
dependencies in Python,
removing these dependencies 
across the source code and configuration files
of a project requires manual effort and expertise.
To tackle this challenge we
introduce \tool,
an end-to-end system to automate
this process.
\tool\ eliminates unused imports
and package declarations
across a variety of file types,
including Python source
and configuration files
such as
\texttt{requirements.txt}
and \texttt{setup.py}.
\tool's modular design makes it
agnostic to the source of dependency
bloat information,
enabling integration with any
detection tool.
Beyond its contribution when
it comes to automation,
\tool\ also incorporates a
novel dynamic analysis component that
improves dependency detection recall.
Our evaluation of \tool's end-to-end effectiveness on
a ground-truth dataset of 37 merged pull requests
from prior work, shows that \tool\ achieves 98.3\%
accuracy in replicating human-made changes.
To show its practical impact,
we run \tool\ on \empirical{\openSourcePackages}
open-source packages,
identifying and trimming
bloated dependencies in \empirical{\projectsWithBloat} of them.
For each case,
we submit a corresponding pull request,
\empirical{\mergedPRs} of which have already been accepted and merged.
\tool\ is available as an open-source project,
encouraging community contributions and further development.

Video demonstration: \url{https://youtu.be/LqTEdOUbJRI}

Code repository: \url{https://github.com/TrimTeam/PyTrim}
\end{abstract}


\section{Introduction}

Python has emerged as one of the most widely adopted
programming languages globally~\cite{cass2024TopLanguages,octoverse2024} 
partly due to its large ecosystem of third-party packages, 
with the Python Package Index (PyPI) --
currently hosting over 654,000 projects~\cite{pypi}.
While these packages accelerate development and foster reuse
they also contribute 
to a notable software engineering challenge:
\emph{dependency bloat}.
As projects evolve, 
obsolete dependencies are often left in the codebase. 
This creates technical debt, 
which in turn leads to 
increased maintenance overhead, 
expanded attack surface
(e.g., via {\tt \_\_init\_\_.py}, which is executed on import), 
and slower pipelines~\cite{cox2019,soto2021comprehensive,azad2023animatedead,ghavamnia2020}.

Consider a real-world example 
from the \texttt{softlayer-python} project~\cite{softlayer-unused-commit}.
In a commit from May 2022,
developers refactored the command-line interface 
to use the \texttt{rich} library,
replacing the older \texttt{prettytable} library.
While all source code usages of \texttt{prettytable} were removed,
the dependency declaration itself 
was left behind 
in four configuration files,
two of them shown in Fig.~\ref{fig:softlayer-example}:
(\ref{softlayer-setup-py}) {\tt setup.py},
which defines package metadata and
installation configuration for
a Python project,
and (\ref{softlayer-requirements-txt})
{\tt tools/requirements.txt},
which lists the Python dependencies
needed for the project's tooling components.

The Python ecosystem provides
numerous open-source tools
that focus on bloat detection.
Some operate at the file level,
identifying unused imports
within individual Python files \cite{autoflake, pylint}.
Others operate at the project level,
looking for unused package dependencies
by comparing configuration files
against the imports used in the source code
\cite{fawltydeps, deptry}.
While valuable for detection,
these tools do not provide
a removal solution.

\point{Contributions} We present \tool, which introduces:
\begin{itemize}
\item An end-to-end pipeline
      for the detection and removal
      of unused dependencies from Python projects,
	  including integration with three existing detection tools.
\item A novel and practical dynamic analysis technique for
      calculating a project's dependencies,
	  which it uses to enhance the static-based approaches
	  of integrated bloat detection tools.
\item Removal logic for dependency declarations
      from a wide array of file formats,
	  including Python source files,
	  \texttt{requirements} files,
	  TOML (Tom's Obvious, Minimal Language)~\cite{toml24}
	  configurations,
      and \texttt{setup.py} files.
\item Automation of the maintenance lifecycle
	  by creating a new Git branch, committing the changes,
      and generating a pull request (PR) with a detailed report,
	  enabling a ``review-and-merge" approach to dependency cleanup.
\end{itemize}

\begin{figure}[t]
\centering
\begin{subfigure}{0.45\linewidth}
\begin{minted}[fontsize=\scriptsize,linenos,xleftmargin=20pt,breaklines]{python}
install_requires = [
  'prettytable',
  ...
  'rich==14.0.0',
]
\end{minted}
\caption{\tt setup.py}
\label{softlayer-setup-py}
\end{subfigure}
\begin{subfigure}{0.45\linewidth}
\begin{minted}[fontsize=\scriptsize,linenos,xleftmargin=20pt,breaklines]{bash}
prettytable
...
rich==14.0.0
\end{minted}
\caption{\tt tools/requirements.txt}
\label{softlayer-requirements-txt}
\end{subfigure}
\caption{
Two of the four configuration files of the {\tt softlayer-python} package.
Its {\tt prettytable} dependency has been unused since commit {\tt 1238377}~\cite{softlayer-unused-commit}.
}
\label{fig:softlayer-example}
\vspace{-7mm}
\end{figure}

\label{sec:background}
\section{Background \& Motivation}
For the remainder of this paper,
we use the term \emph{configuration files}
to refer to auxiliary files
that declare a package's dependencies
or define its installation behavior.
Such files include {\tt requirements.txt},
{\tt pyproject.toml}, and {\tt setup.py}.
Accordingly,
we use the term \emph{source files}
to refer to source code files
that implement
the main functionality of a package.
Finally,
we use the term \emph{project}
to refer to the source code
(usually published on GitHub)
of a corresponding Python \emph{package},
which is in turn published on PyPI.

We define a dependency package
as \emph{unused} (or \emph{bloat})
if it is declared in the project's configuration files
but is not used anywhere in the project
beyond, at most, a direct import.
While we acknowledge that
importing a package causes Python
to implicitly execute
its {\tt \_\_init\_\_.py} file~\cite{import-python-docs},
which could
be part of the program logic,
we encounter no such cases during our evaluation.

To detect unused dependencies such as the
{\tt prettytable} dependency in {\tt softlayer-python}
from Fig.~\ref{fig:softlayer-example},
developers must \textit{run}
a dependency bloat detection tool,
such as {\tt FawltyDeps}~\cite{fawltydeps},
on the package's source code and
manually \textit{examine} the tool's output
to identify bloated dependencies.
Note that the process of detecting unused dependencies
comprises two distinct phases:
(i) computing the set of the package's dependencies,
and (ii) determining the extent to which each of them is used.

Once unused dependencies are identified,
developers must undertake a manual removal
process that involves
\textit{removing} dependency declarations from configuration files
and imports from Python files,
and finally \textit{submitting} a PR
to fix the problem.
This remediation workflow requires careful attention
to ensure that all references to the unused
dependency are properly eliminated from the
codebase without affecting the package's functionality.

\point{Challenge 1: Project dependency resolution}
To compute the dependency set,
existing detection tools rely on
static analysis of configuration files.
While tools such as {\tt FawltyDeps},
correctly parse standard configuration
files such as the ones depicted
in Fig.~\ref{fig:softlayer-example},
they fail when dependencies are programmatically constructed.
For example, in the {\tt optimizely-sdk} package
(Fig.~\ref{fig:optimizely-sdk}),
the {\tt setup.py} file dynamically builds
the {\tt install\_requires} list by reading an external file,
making static analysis approximation difficult
and causing missed dependencies.
Given that over 50\% of PyPI packages published in
April 2024~\cite{pypi-stats} use {\tt setup.py}
as their configuration file,
the practical approximation of its behavior
is essential.

\point{Challenge 2: Multitude of
configuration files and types}
To remove dependency declarations
developers have to search for
and edit all related configuration files
(e.g., all four aforementioned configuration
files in the case of {\tt softlayer-python}).
Notably,
open-source projects have previously
encountered issues~\cite{pvanalytics-configuration-skew}
with partial updates of configuration files,
leading to broken builds.

\begin{figure}[t]
\centering
\begin{subfigure}{0.49\linewidth}
\begin{minted}[fontsize=\scriptsize,linenos,xleftmargin=20pt,breaklines]{python}
f = open('reqs/core.txt')
REQS = f.read().splitlines()
...
install_requires = REQS
\end{minted}
\caption{\tt setup.py}
\label{optimizely-setup-py}
\end{subfigure}
\begin{subfigure}{0.49\linewidth}
\begin{minted}[fontsize=\scriptsize,linenos,xleftmargin=20pt,breaklines]{bash}
pyrsistent>=0.16.0
...
\end{minted}
\caption{\tt reqs/core.txt}
\label{optimizely-requirements-txt}
\end{subfigure}
\caption{
{\tt optimizely-sdk} builds its dependencies dynamically,
causing static analysis-based detectors
to miss its {\tt pyrsistent} dependency
and fail to identify it as unused.
}
\vspace{-3mm}
\label{fig:optimizely-sdk}
\end{figure}

\point{Challenge 3: Human error}
While valuable for detection,
existing tools do not provide a removal solution.
That is,
developers have to perform the
remediation workflow manually,
making the procedure prone to human error.
This gap between
detection and remediation
represents a bottleneck
in automated software maintenance.

\section{PyTrim}

\subsection{Overview}

\tool\ receives as input
the path to a Python project
and then proceeds to
(a) dynamically infer
    the project's dependencies
	by installing it in isolation,
(b) identify the subset
    of unused dependencies
	by invoking an unused dependency detector, and
(c) locate and remove their occurrences
    (declarations and imports)
	across a wide range of file formats.
Its design is detector-agnostic,
allowing any dependency analysis tool to be integrated.

\subsection{System Architecture}

\begin{figure}[t]
  \centering
  \includegraphics[width=\linewidth]{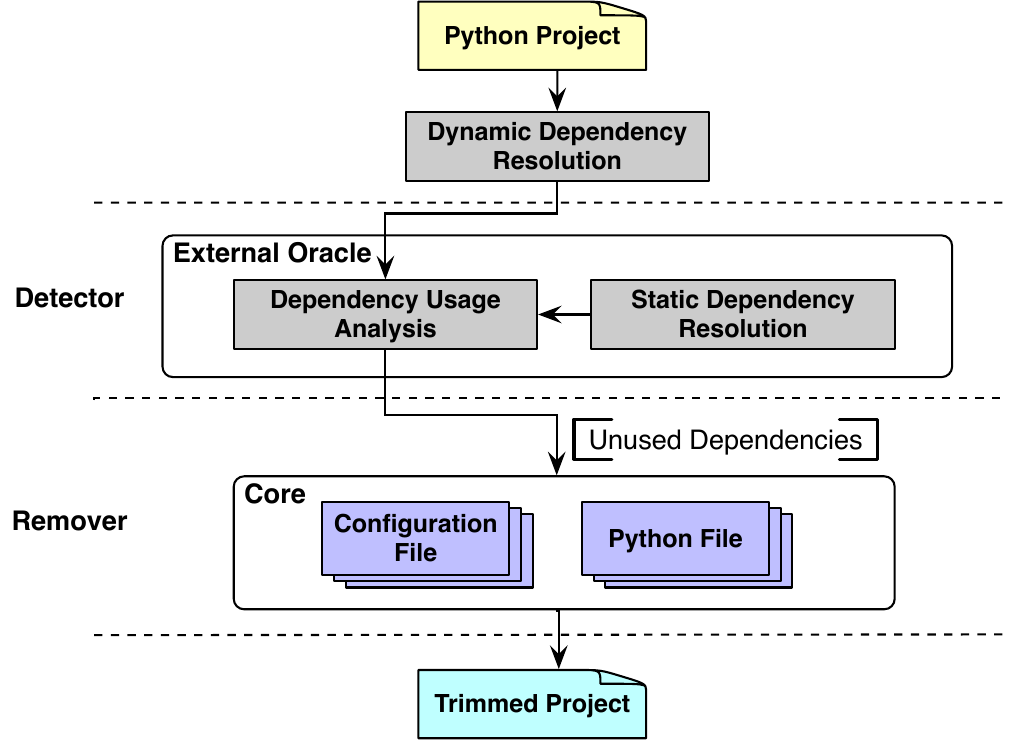}
  \caption{High-level architecture of \tool.}
  \label{fig:pytrim_architecture}
\vspace{-5mm}
\end{figure}

\tool~incorporates three primary modules:
\begin{itemize}
    \item \textbf{Dynamic dependency resolver:}
	A component that generates a list of dependencies
	based on install-time information.
    \item \textbf{Detector:}
    An external dependency analysis tool
	(e.g., extended PyCG~\cite{drosos2024bloat})
    to determine which dependencies are unused.
    \item \textbf{Remover:}
    A component that implements file-specific logic
    to locate and remove unused dependency occurrences
	(declarations and imports).
\end{itemize}
The interaction between these modules and the overall workflow
is shown in Fig.~\ref{fig:pytrim_architecture}.

\point{Dynamic dependency resolver}
To address Challenge 1,
we perform a source installation
of the project under analysis
by invoking {\tt pip}
with the ``{\tt -t}" argument,
to install artifacts in an isolated environment.
This is a dynamic operation,
meaning that it yields no false positives,
while at the same time requires no inputs.
This step is inspired by recent work
that has been introduced to support
cross-language call-graph
construction~\cite{pyxray}.
Then,
we utilize
the {\tt pipdeptree}~\cite{pipdeptree} utility
to produce a directed graph 
representation of installed packages
in this environment,
where an edge from {\tt x} to {\tt y}
denotes that {\tt x} depends on {\tt y}.
The package under analysis is the vertex
with no incoming edges,
and its dependencies are the destinations
of its outgoing edges.
In this manner,
we dynamically obtain a list of the project's dependencies.
A current limitation of our approach
is that it only considers
the default installation options,
and thus only follows one concrete execution path
in the configuration file.

\point{Detector}
As a default,
\tool\ utilizes a call-graph based
detector~\cite{drosos2024bloat}.
It also provides out-of-the-box support 
for two other popular detectors,
namely {\tt FawltyDeps}~\cite{fawltydeps}
and {\tt deptry}~\cite{deptry}.
We modify each detector's
dependency resolution component
to take into account
the union of its 
built-in static dependency resolver
and \tool's dynamic dependency resolver,
to decrease missed dependencies.

\point{Remover}
To address Challenge 2 we follow a two-fold approach.
For Python source files (\texttt{.py}),
\tool\ employs Abstract Syntax Tree (AST) analysis
to identify and remove unused imports.
For configuration files,
it uses a combination of methods based on file type:
structured file types such as \texttt{TOML}, \texttt{YAML},
and \texttt{INI/CFG} are managed by
parsing libraries such as \texttt{toml},
\texttt{PyYAML}, and \texttt{configparser};
line-based formats including \texttt{requirements.txt}
and their common \texttt{.in} counterparts
are handled with regular expressions;
and executable \texttt{setup.py}
files are processed via AST analysis.

To ensure dependency integrity 
after these modifications,
\tool\ identifies 
when a lock file (e.g., \texttt{poetry.lock}) 
may be out of sync
and prints a message 
to inform the user 
to regenerate it manually,
as automatic regeneration 
could be unsafe 
due to project-specific versioning or hashes.
Similarly, 
less structured files 
like shell scripts or Dockerfiles,
are only analyzed for reporting
and are not modified automatically 
due to their inherent complexity and risk
of reliably altering their syntax programmatically.

\subsection{Implementation and Usage}
\tool\ is implemented in \linesOfCode{} lines of Python code
and is provided as a command-line tool,
installable from PyPI via \texttt{pip install pytrim} and
accessible via the \texttt{pytrim} command.
Apart from end-to-end automated analysis
(Challenge 3),
it can also work in removal-only mode,
bypassing its built-in detectors,
accepting the list of
unused packages as input alongside the project's path.
Further,
\tool\ can also generate markdown reports and PRs.

\section{Evaluation}
\subsection{Dynamic Dependency Resolution}
We compare \tool's dependency resolver
with the static resolvers employed by
the three integrated detection tools,
to quantify the missed dependencies it helps uncover.
To this end,
we evaluate each resolver
on a dataset of 1,300 popular GitHub projects
studied in previous work~\cite{drosos2024bloat}.
From these, we only keep
a subset of 971 packages
that are successfully installed.
Compared to the static resolvers,
the dynamic resolver
is able to uncover
missed dependencies
in 48/971 (5\%) of cases.
The main causes for the static
resolvers' misses are:
(a) parsing limitations for
edge-cases of supported declarative files,
such as {\tt pyproject.toml} (18 cases),
(b) arbitrary {\tt setup.py} code,
such as function invocations (22 cases),
and (c) miscellaneous bugs (8 cases).
For example,
the dependency of
{\tt optimizely-sdk} on {\tt pyrsistent},
shown in Fig.~\ref{fig:optimizely-sdk},
is detected only
through our dynamic resolution technique.

\subsection{Remover effectiveness}
To validate \tool's effectiveness, 
we design an automated pipeline 
to measure its accuracy 
against real-world software maintenance tasks.
As a ground truth,
we use the dataset from
Drosos et al.~\cite{drosos2024bloat},
which provides a curated set of PRs
where bloated dependencies 
were manually removed from open-source Python projects.

The ground truth for each PR
consists of two parts:
the list of dependencies
the developer removed,
and the final,
correct state,
of the modified files.
Our pipeline 
first runs \tool\ on the pre-PR version of each project,
providing the ground-truth list of dependencies as input.
It then 
compares the files modified
by \tool\ against the ground-truth files 
to measure replication accuracy,
ignoring non-semantic differences 
such as changes in white space or comments.

\begin{table}[t]
\centering
\caption{\tool's replication accuracy on manually
created PRs from previous work~\cite{drosos2024bloat}.}
\label{tab:evaluation_results}
\begin{tabular}{lr}
\toprule
\textbf{Metric} & \textbf{Value} \\
\midrule
Total Pull Requests Analyzed & 37 \\
Total Files with Dependency Changes & 76 \\
Files Excluded (e.g., Documentation) & 16 \\
\textbf{Relevant Files for Comparison} & \textbf{60} \\
\midrule
Files Correctly Replicated by \tool & 59 \\
Files with Mismatched Output & 1 \\
\textbf{Replication Accuracy} & \textbf{98.33\%} \\
\bottomrule
\end{tabular}
\vspace{-3mm}
\end{table}

The results of our analysis are summarized in Table~\ref{tab:evaluation_results}.
Out of 60 relevant files where dependencies were removed,
\tool\ successfully matches the final state of 59,
achieving a replication accuracy of 98.33\%.
It is also efficient;
when provided with the pre-computed list of unused dependencies,
it processes the entire evaluation dataset of 37 projects
in less than 10 seconds.

The only mismatch occurred
in a PR~\cite{drosos-pr}
on {\tt simple-salesforce},
revealing a limitation of \tool's scope:
dependency refactoring.
While \tool\ correctly removed
\texttt{cryptography},
the human developer also
refactored \texttt{pyjwt}
into \texttt{pyjwt[crypto]},
to include optional cryptographic features.
Such semantic changes
require domain-specific knowledge
and
are currently 
out of scope for \tool,
which focuses solely
on removing unused dependencies,
not refactoring them.

\subsection{Real-World Deployment}
To evaluate \tool's practical applicability,
we deploy it 
on the 971 installable open-source Python projects
studied in previous work~\cite{drosos2024bloat}.
For each project, 
we run \tool~using
the call-graph-based detector
to identify unused dependencies.
Then, 
we manually verify all changes
and submit PRs
to the respective repositories.

\tool\ removes unused dependencies
in \projectsWithBloat{} projects,
resulting in \projectsWithBloat{}
submitted PRs.
At the time of writing,
\mergedPRs{} PRs have been merged 
by the project maintainers, 
\pendingPRs{} are still under review, 
and \closedPRs{} were closed
due to project-specific policies.

As an example consider
the \texttt{softlayer-python} project
(Fig.~\ref{fig:softlayer-example}).
\tool\ correctly identified 
that the \texttt{prettytable} dependency
is unused.
Then,
it generated a PR 
to remove the dependency 
from four configuration files
(\texttt{setup.py} and
three separate \texttt{requirements.txt} files).
\tool\ also identified
the unused package name
in the \texttt{README.rst} file
and
flagged it for manual review.
The resulting PR, 
with the manual edit for the README,
was merged 
by the project maintainers.
Merged PRs to the 3 most popular projects
are shown in Table~\ref{tab:merged_prs}.
The complete list is available in our GitHub repository.

Notably,
bloated dependencies in
{\tt edx/edx-lint} and
{\tt optimizely/python-sdk}
(Fig.~\ref{fig:optimizely-sdk})
would have gone undetected without \tool,
as existing detectors
fail to accurately identify
their dependencies.

\begin{table}[t]
\centering
\caption{Merged PRs to the most popular projects (by stars).}
\label{tab:merged_prs}
\begin{tabular}{@{}llr@{}}
\toprule
\textbf{Project} & \textbf{PR ID} & \textbf{Stars} \\
\midrule
\texttt{napalm-automation/napalm} & \texttt{\#2243} & 2,374 \\
\texttt{pycqa/prospector} & \texttt{\#781} & 2,035 \\
\texttt{django-json-api/django-rest-framework-json-api} & \texttt{\#1285} & 1,238 \\
\bottomrule
\end{tabular}
\vspace{-5mm}
\end{table}

\section{Related Work}

While software debloating is a broad research area,
this section focuses on work most directly related
to Python dependency analysis.
For comprehensive surveys of debloating techniques in other ecosystems,
we refer readers to Drosos et al.~\cite{drosos2024bloat}
and the Systematization of Knowledge (SoK) work
by Alhanahnah et al.~\cite{alhanahnah2024sok}.
Most relevant to our work is Drosos et al.~\cite{drosos2024bloat},
who conducted a large-scale study on dependency
bloat in the Python ecosystem and established a
methodology for identifying unused dependencies
through fine-grained call graph analysis using
the PyCG static call graph generator~\cite{salis2021pycg}.
This research provides the foundational understanding
of the problem space that \tool\ builds upon.

Existing Python tools
focus primarily on detection
and
can be broadly categorized into two groups.
The first consists of source code linters,
such as
\texttt{autoflake}~\cite{autoflake} and
\texttt{pylint}~\cite{pylint},
that identify
unused imports
directly within \texttt{.py} files.
While some of these,
such as \texttt{autoflake},
can remove the identified import lines,
their scope
is limited to the source code;
they do not modify project-level
configuration files.
The second group operates at the project level;
tools such as
\texttt{deptry}~\cite{deptry} and
\texttt{FawltyDeps}~\cite{fawltydeps}
compare declared dependencies
against source code imports
to find unused packages,
but they do not automate
the removal of the corresponding
\texttt{import} statements.

\section{Conclusion}

We presented \tool,
a system for end-to-end removal
of unused dependencies
and imports.
We outlined its
detector-agnostic architecture,
its novel dynamic dependency resolution technique,
and
its support for a wide range
of configuration formats.
Our evaluation demonstrates
its effectiveness
in uncovering
previously undetectable
dependency bloat
and
automatically removing it.

\bibliographystyle{IEEEtran}
\bibliography{pytrim}

\end{document}